\documentstyle[prd,aps,floats]{revtex}
\draft
\newcommand{\lsim}{\mathrel{\mathop{\kern 0pt \rlap
  {\raise.2ex\hbox{$<$}}}
  \lower.9ex\hbox{\kern-.190em $\sim$}}}
\newcommand{\be}{\begin{equation}}
\newcommand{\ee}{\end{equation}}
\newcommand{\ba}{\begin{eqnarray}}
\newcommand{\ea}{\end{eqnarray}}
\newcommand{\nn}{\nonumber}
\newcommand{\ep}{\epsilon}

\newcommand{\npb}{Nucl. Phys. B}
\begin{document}
\twocolumn[\hsize\textwidth\columnwidth\hsize\csname
@twocolumnfalse\endcsname
\title{
\hbox to\hsize{\large Submitted to Phys.~Rev.~D \hfil E-print: hep-ph/980000}
\vskip1.55cm
Finite temperature effects on baryon transport \\
scattering in the early Universe
}

\author{
In-Saeng Suh$^{\rm a,b}$ 
and G. J. Mathews$^{\rm a}$
}

\vspace{0.8 cm}

\address{
\begin{tabular}{c}
$^{\mbox{a}}$
Department of Physics,
University of Notre Dame,\\
Notre Dame, IN 46556, USA 
\vspace{0.3 cm}
\\
$^{\mbox{b}}$
The Research Institute for Natural Sciences, \\ 
Hanyang University, 133-791, Seoul, Korea
\end{tabular}
}

\date{\today}
\maketitle   
 
\begin{abstract}
We have computed finite temperature corrections to the electron-hadron scattering
cross sections. These are based upon the renormalized 
electron mass and the modified density of states
due to the presence of a background thermal bath. 
It is found that the electron-hadron thermal 
transport scattering cross section can be much larger than the 
zero temperature one. 
In the case of electron-neutron transport scattering, we find
$\sigma_{ne}(T) / \sigma_{ne} (T=0) \simeq 5$ at $T \simeq 0.1 \, MeV$.

\vskip 0.3cm
\end{abstract}

\vskip2.2pc
]
 
\narrowtext
\newpage
\section{Introduction}

Finite temperature effects on elementary processes are significant from
the point of view of cosmology and astrophysics. The early universe
is usually described as a hot gas of particles in nearly thermodynamical
equilibrium. Temperature effects
enter through the statistical distribution functions. These can renormalize
the masses and the wave functions. 
These renormalized masses and wave functions can then affect scattering processes
and decay rates.
Several authors \cite{donoghue} have generalized
the electron-mass and wave-function renormalization to all temperatures
and densities. 
Dicus et al. \cite{dicus} and independently, Cambier et al. \cite{cps} included 
the finite temperature effects on weak reaction rates in calculations of 
standard big-bang nucleosynthesis $(BBN)$. 
They obtained the corrected light-element 
abundance and found that the corrections are only of order of a few percent. 
After that, Saleem \cite{saleem} included the effects of the electron mass shift 
at finite temperature on $BBN$ and Baier et al. \cite{baier} examined the finite 
temperature radiative corrections to the weak neutron-proton decay rates. 
More recently, Fornengo et al. \cite{jwkim} have considered the finite temperature 
effects on the neutrino decoupling temperature which is important in the evolution 
of the early universe. In the present work we consider finite temperature corrections
to electron-hadron scattering which is important for baryon inhomogeneous cosmologies.  

Baryon inhomogeneities might have been produced during the cosmological quark-hadron
phase transition in the early universe \cite{witten}.
If such inhomogeneities were present, then the different diffusion
lengths for neutrons and protons could lead to the formation of
high-baryon density proton-rich regions and low-baryon density neutron-rich regions.
The light element nucleosynthesis yields from such regions can differ significantly 
from those of standard homogeneous big-bang nucleosynthesis \cite{AHS}.
In view of the importance of using light-element yields from the $BBN$ to 
constrain the baryon-to-photon ratio as well as various cosmological and 
particle physics theories, such inhomogeneous models must be examined 
seriously. It is therefore important to quantify the effects of baryon diffusion 
as accurately as possible. 

In this regard Applegate, Hogan and Scherrer (AHS) \cite{AHS} have calculated the
diffusion rate of baryons through the electron-positron plasma in the early universe. 
Subsequently, several authors used their results in calculations of 
inhomogeneous $BBN$ \cite{applegate,mathews}.
In $AHS$ it was suggested that the diffusion coefficients could be derived from the 
mobility of the heavy particles, and that the mobility is determined from the distribution
functions of the background plasma and the transport cross section.
It is important, therefore, to carefully quantify the values of the 
distribution functions and the transport cross sections.

However, in all previous baryon diffusion coefficient calculations,
vacuum transport scattering cross sections have been used.
Therefore, in order to estimate the baryon diffusion coefficients more precisely,
in the present work we take into account the finite temperature effects in the
calculation of baryon diffusion coefficients at temperatures $\lsim \; MeV$.

Specifically, we calculate the transport scattering cross section of elastic
electron-hadron scattering at finite temperature. 
Here we shall treat hadrons as particles which have an internal structure and 
an anomalous magnetic moment (although we do not have a good field theory for 
the magnetic moments of protons or neutrons at finite temperature).
Also, we assume that their internal structure is not affected by finite
temperature since their mass is more than about 1000 times the temperatures of interest.

The plan of the paper is as follows. In Section \ref{sec:finite}, we discuss 
how to include finite temperature effects in the calculation. 
In particular, we will briefly discuss the effective mass of an electron 
in the MeV temperature range.
In Section \ref{sec:scattering}, we evaluate the electron-hadron transport 
scattering cross section at finite temperature. 
Finally, we summarize our results and discuss some astrophysical applications.
We shall employ units in which $\hbar = k_B = c = 1$, except when specific
units must be attached to a result.

\section{Finite temperature effects}
\label{sec:finite}

In the early universe, where the particles propagate in 
a thermal bath rather than in vacuum, their dynamics and 
interactions are modified to some extent.
The behavior of particles in a thermal bath is systematically 
described in the framework of finite temperature quantum field theory
\cite{FTQFT}.
In a finite temperature analysis, the following points must be taken into 
account \cite{donoghue}:

1) The free spinors used in the derivation of the cross section
must be replaced by free finite temperature spinors $u_{T}(p)$ which
describe freely propagating particles in the background thermal bath.
Also, the propagators are modified even at the tree level. Since these break
the Lorentz invariance, they are absorbed into the effective mass 
for the particle. This effect can be evaluated by calculating the self--energy
of the particle in the heat bath. 
At the same time, it modifies the cross section due to the change of propagators and 
spinors in the scattering amplitude and the distribution functions $f(E)$;

2) The density of final states used in the determination of the cross section
is modified by the background thermal bath as follows:
\begin{equation}
\frac{{\rm d}^3 p\,'}{(2 \pi)^3 2 E'}
\longrightarrow
\frac{{\rm d}^3 p\,'}{(2 \pi)^3 2 E'} [1 - f_F(E')],
\end{equation}
where $f_F(E')$ is the Fermi-Dirac distribution at temperature $T$ for final energy
$E'$.
This takes into account the states already occupied by fermions in the thermal
bath.
However, note that since we are still interested in single particle scattering,
we do not make a thermal average over initial states. 
 
The dynamics of electrons in a thermal bath is 
modified by the electromagnetic interactions with background photons
and electrons themselves. Therefore, the effect of the thermal bath 
on the propagation of an electron is expressed by calculating
the electron self--energy in the presence of the ambient $e^+$, $e^-$
and $\gamma$'s \cite{donoghue}. 
The temperature corrected electron physical mass is then
obtained by evaluating the renormalized propagator and finding the zero of its
inverse. Thereby, the temperature-dependent physical mass of the electron 
is given as \cite{donoghue}
\ba
m_{T}^2 &\equiv& m^2 = E^2-\vec{p}\,^2 \nonumber \\
&=&
m_0^2 + \frac{2}{3}\pi\alpha T^2 +  \frac{4}{\pi} \alpha T^2 {\cal B}(x) + 
\frac{m_{0}^2}{2 \pi^2} \alpha {\cal J}(p) ,
\ea
where $m_0 = 0.511 \, MeV$ is the electron rest mass in vacuum.  
The function ${\cal B}(x)$ with $x = T/m_0$, is defined as \cite{donoghue}
\be
{\cal B}(x) \equiv \frac{1}{x} \int_{1/x}^{\infty} 
ds \,\frac{\sqrt{s^2 x^2- 1}}{e^{s} + 1} ,
\ee
and
\ba
{\cal J}(p) = \int \frac{d^3 k}{E_k} f_F(E_k) 
&[& \frac{1}{ E_p E_k + m_{0}^{2} - \vec{p} \cdot \vec{k}} \nn \\
&-&  \frac{1}{ E_p E_k - m_{0}^{2} + \vec{p} \cdot \vec{k}} \; ] ,
\ea
where $E_{k}^{2} = m_{0}^2 + k^2$. In Ref. \cite{jwkim} it is has been shown that 
the ${\cal J}(p)$ 
term is negligible for $T \sim MeV$. We will therefore neglect this
term in our analysis.
 
Eq. (2) is valid for all temperatures. 
It gives the correct result $m_{T} = m_0$ at $T=0$. 
Around $T \sim m_0$, however,
the third term becomes important and has to be taken into
account [for example, ${\cal B}(x=1) \simeq 0.543$].
It also has been shown in \cite{jwkim} that the thermal corrections to the
electron mass at $T\sim $ MeV are sizeable. At $T=1$ MeV
the electron mass increases by 4.1\%; and at $T=2$ MeV the correction
is as large as 16\%. 
The change in effective mass of a particle modifies its
contribution to the energy density of the universe, and therefore 
the expansion rate.
The modification of the electron mass
also changes the relationship between the neutrino and 
photon temperature.
At finite temperature, the photon propagator has an additional term 
proportional to the photon phase-space density \cite{dicus}.

In brief,
the influence of finite temperature on the particle evolution
in the early universe is due to the temperature--dependent shifts in
the dynamical mass of the particles and the
temperature--dependent modification of the interactions between particles.
Both modifications will be included in the calculation of the scattering
cross section described in Section \ref{sec:scattering}.

\section{Electron--hadron transport scattering at finite temperature} 
\label{sec:scattering}

In this section we calculate the transport cross section of electron--hadron
scattering at finite temperature which is one of the fundamental processes 
in cosmology and astrophysics.
In 1950, Rosenbluth \cite{rosenbluth} calculated the electron-hadron 
differential cross section (the so called Rosenbluth formula) under the assumption 
that the electron is ultrarelativistic($m \ll E$) and 
in the rest frame of  the incoming hadron. This treatment takes into account the 
internal structure and anomalous magnetic moment. 
Applegate, Hogan and Scherrer \cite{AHS} used the Rosenbluth formula in the 
calculation of their neutron diffusion coefficient
for electron-neutron scattering with the assumption that the electron energy 
is much less than neutron mass $M$. They obtained a constant transport
cross section $\sigma_{t}$ for the vacuum interaction between an electron and a 
neutron \cite{AHS},
\be
\sigma_{t}^{AHS} = 3 \pi \left(\frac{\alpha \kappa_n}{M}\right)^2 
\simeq 8 \times 10^{-31} \; cm^2
\ee
where $\alpha=e^2 /\hbar c$ is the electron fine structure constant and $\kappa_n$
is the anomalous magnetic moment of the neutron.

However, at $T \sim MeV$,
the dynamical properties of electrons and photons in this thermal bath would be 
changed (though hadrons would not be affected by the background thermal bath since
their masses are are more than about 1000 times the mass of the electron). 
We therefore only have to take into account the effect of finite temperature on the
interaction between electrons and hadrons.    

The transport cross section $\sigma_t (T)$ of the process
\be
e^- (E,\vec{p}) + H(\epsilon, \vec{k}) 
\longleftrightarrow  e^- (E',\vec{p}') + H(\epsilon', \vec{k}') \;,
\ee
where $H$ denotes a hadron, is defined by 
\be
\sigma_t (T) = \int d \bar{\sigma} (1 - cos\theta'),
\ee
where $\theta'$ is the scattering angle.  The differential cross section 
(including the thermal phase space) is 
\ba
d \bar{\sigma} &=& \frac{m}{E} \frac{M}{\epsilon} \frac{1}{|\vec{J}_{inc}|V}
 \frac{m}{E'} \frac{{\rm d}^3 p'}{(2\pi)^3}
 \frac{M}{\ep'} \frac{{\rm d}^3 k'}{(2\pi)^3} {\cal S}(E', \ep') \nn \\
&~& \times \; (2\pi)^4 \delta^{(4)}(p+k-p'-k') \bar{|{\cal M}|^2} \;, 
\ea
where $m$ is the mass of electron, $M$ is the mass of hadron, 
$V$ is the normalization volume, and the statistical 
factor ${\cal S}(E',\ep')$ is
\be
{\cal S}(E',\ep') = [1 - f(E')][1 - f(\ep')],
\ee
where $f(E)$ is the Fermi-Dirac distribution function.
The flux is given by the number of particles passing through a unit area 
per unit time,
\[
|\vec{J}_{inc}| = \frac{|\vec{v} - \vec{V}|}{V}, \;\;\;\;\;\;\;\;\; 
\vec{v} = \frac{\vec{p}}{E}, \;\;\;\;\; \vec{V}=\frac{\vec{k}}{\ep} \, ,
\]
where $\vec{v}$ and $\vec{V}$ denote the initial velocities of electrons and hadron,
respectively.
The square of the spin-averaged scattering matrix element is given by
\ba
\bar{|{\cal M}|^2} &=& \frac{1}{4} \sum_{spins} |\bar{u}(p', s') \gamma^{\mu} u(p,s)
[\alpha  {\cal D}_F (q^2)] \nn \\
&~& \;\;\;\;\;\;\;\;\; \times \; \bar{u}(k', S') \Gamma_{\mu}(k', k) u(k,S)|^2  \\
&=& \alpha^2 |{\cal D}_F (q^2)|^2 {\cal L}^{\mu \nu}{\cal H}_{\mu \nu} \, ,
\ea
where $s$ and $S$ denote the electron and hadron spin respectively.  The vertex
function $\Gamma_{\mu}(k', k)$ is defined using the Gordon decomposition is \cite{qed}
\be
\Gamma_{\mu}(k', k) =  \gamma_{\mu}(F_1 (q^2) + F_2 (q^2)) 
- \frac{F_2 (q^2)}{2M}(k'+k)_{\mu} .
\ee
Here, $q = k' - k$, is the momentum transfer and $F_1 (q^2)$ and $F_2 (q^2)$ are
hadron form factors.
The $q^2 \rightarrow 0$ limit of the proton and neutron form factors 
are known from scattering experiments \cite{qed}:
\ba
\left. \begin{array}{cc}
F_{1}^{p}(0) = 1, & F_{1}^{n}(0) = 0, \\ 
F_{2}^{p}(0) \equiv \kappa_p = 1.792, & \;\;\;\;\;\;\; 
F_{2}^{n}(0) \equiv \kappa_n = -1.913 .
\end{array}  \right.
\ea
The lepton tensor in Eq. (11) is
\[
{\cal L}^{\mu \nu} = \frac{1}{2 m^2} [p'^{\mu} p^{\nu} + p^{\mu} p'^{\nu} 
- g^{\mu \mu}(p' \cdot p - m^2)],
\]
and the hadron tensor is
\ba
{\cal H}_{\mu \nu} &=&  \frac{1}{2 M^2} \{ (F_1 + F_2)^2
[k_{\mu} k'_{\nu} + k'_{\mu} k_{\nu} - g^{\mu \mu}( k' \cdot k - M^2)] \nn \\
&~& - \left[(F_1 + F_2) F_2 - \frac{1}{4} F_{2}^{2} \left(1+ \frac{k' \cdot
k}{M^2}\right) \right](k'_{\mu} +k_{\mu}) \nn \\
&~& \times \; (k'_{\nu}+k_{\nu}) \}. 
\ea
The finite temperature parts of the photon propagator give additional
radiative corrections to the Feynman diagrams involving virtual photons.
The photon propagator in the Landau gauge is \cite{donoghue}
\be
{\cal D}_F (q^2) = - 4 \pi \left[\frac{i}{q^2 + i \ep} + 2 \pi f_{B}(\omega) \delta(q^2)
\right],
\ee
where the phase-space density $f_B (\omega)$ is the Bose-Einstein distribution function
for photons.
[The factor of $- 4 \pi$ arises from our use of  Gaussian units;
in ``rationalized'' units, this factor in Eq. (15) is replaced by $-1$]

In the rest frame of the incoming hadron, $k = (k^{0} = \ep = M, \vec{k} = 0)$.
Integrating Eq. (7) over scattering angle,
we obtain the temperature dependent transport cross section
\be
\sigma_t (T) = \frac{1}{2 \pi} \frac{m^2 M}{|\vec{p}|^2} \int_{E'_{min}}^{E'_{max}} 
d E' (1 - \beta(E'))
\bar{|{\cal M}|^2} {\cal S}(E',\ep') \, ,
\ee
where
\be
\beta(E') = \frac{E'(M+E) - m^2 - M E}{|\vec{p}||\vec{p}'|} .
\ee
For a given initial electron energy $E$, the kinematical limits for the final-state
energy of the electron, $E'_{min}$ and $E'_{max}$, can be determined from the constraint
$|cos \theta'| \leq 1$.
The squared spin-averaged scattering matrix element in Eq. (16) becomes
\be
\bar{|{\cal M}|^2} = 2 \left(\frac{\pi \alpha}{m M}\right)^2 {\cal F}(E') ,
\ee
where
\ba
{\cal F}(E') &=& \left(\frac{2 E E'+M(E'-E)}{(E'-E)^2}\right)
\left[F_{1}^{2} - \frac{1}{2} F_{2}^{2}\left(\frac{E'-E}{M}\right)\right] \nn \\
&~& +\;(F_1 + F_2)^2 \left[1 - \frac{m^2}{M(E'-E)}\right] .
\ea
Since $q^2 = 2 M (E'-E)$ in the hadron rest frame, the second term in 
Eq.(15) does not contribute in the calculated transport scattering cross section
(because $\beta(E) = 1$ for $E' = E$).
From Eqs. (16) and (18), we finally find that the electron-hadron transport cross 
section at finite temperature is
\be
\sigma_{t} (T) = \pi \alpha^2 \frac{1}{M |\vec{p}|^2}
\int_{E'_{min}}^{E'_{max}}
d E' (1 - \beta(E')) {\cal F}(E') {\cal S}(E',\ep') \, ,
\ee   
where $|\vec{p}|^2 = E^2 - m^2$.

For  neutron scattering, $F_1 = 0$ and 
$F_2 = \kappa_n$. Thus, Eq. (19) becomes 
\be
{\cal F}(E') = \kappa_{n}^{2} \left(\frac{1}{2} - \frac{E E' + m^2 }{M (E'-E)}\right) .
\ee
In the case of ultrarelativistic $(UR)$ electrons in which the electron mass 
can be ignored 
($m \ll E$),
the electron-neutron transport cross section at finite temperature is given by
\be
\frac{\sigma_{ne}^{UR} (T)}{\sigma_{t}^{AHS}} =  \frac{M^2}{6 E^2} 
\int_{E'_{min}}^{E'_{max}} dE'\left(\frac{E-E'}{E \, E'} + \frac{2}{M}\right) 
{\cal S}(E',\ep').
\ee
Using a Taylor series expansion,
at zero temperature and for $E \ll M$, analytically we can obtain 
\be
\frac{\sigma_{ne}^{UR} (T)}{\sigma_{t}^{AHS}} \simeq 1.
\ee
We can see this result in Fig. 2 for the numerically evaluated 
$\sigma_{ne}^{UR} (T)$. 

For the scattering of an electron with a proton, Eqs. (13) and (18) yield 
\ba
{\cal F}(E') &=& \left(\frac{2 EE'+M(E'-E)}{(E'-E)^2}\right) 
\left[1 - \kappa_{p}^{2} \left( \frac{E'-E}{2 M}\right) \right] \nn \\
&~& + \; (1 + \kappa_p)^2  \left[1 - \frac{m^2}{M(E'-E)}\right].  
\ea
In the case of ultrarelativistic electrons, Eq. (24) becomes
\ba
{\cal F}(E') &=& \frac{2 EE'+M(E'-E)}{(E'-E)^2} \nn \\ 
&~& + \; (1 + \kappa_p)^2  
- \kappa_{p}^{2} \left[\frac{1}{2} + \frac{E E'}{M (E'-E)} \right].  
\ea
For electron-proton collisions, the most important scattering
mechanism is Coulomb scattering and the differential cross section is given by the 
Mott formula. In which case the Coulomb transport cross section is \cite{qed},
\be
\sigma_{pe}^{Coul} (T) = 4 \pi \alpha^{2} \left( \frac{E}{|\vec{p}|^2} \right)^2
\Lambda(T) \; [1 - f(E)],
\ee
where $|\vec{p}|$ is the electron momentum. Because Eq. (26) diverges at small angles, 
the usual approximation is to cut-off the angular integration at an angle given
by the ratio of the Debye shielding length, $\lambda_D = (T/e^2 n_e)^{1/2}$, 
to the thermal wave length, $\lambda_{th} = (2 \pi  / m T)^{1/2}$ \cite{banerjee}.
This defines to  Coulomb logarithm $\Lambda (T) = ln (\lambda_D / \lambda_{th})$. 
With this we can evaluate numerically the transport cross sections for a given initial
electron energy.

In the early universe, the number density and
energy density of electrons are given by \cite{kolb}
\ba
n_e (T) &=& g_e \int \frac{d^3 p}{(2 \pi)^3} f(E), \\
\rho_e (T) &=& g_e \int \frac{d^3 p}{(2 \pi)^3} E f(E) \,,
\ea
where $g_e$ is the electron degeneracy.
For an electron in thermal equilibrium, the phase space occupancy $f(E)$ is given by 
the Fermi-Dirac distribution
\be
f(E) = \frac{1}{e^{(E-\mu)/T} + 1} \, ,
\ee
where $\mu$ is the electron chemical potential. In the early universe, 
$\mu/T \lsim 10^{-9}$ \cite{dicus}. Therefore, density effects which are
parameterized
by $\mu$ are much less significant than temperature effects which are parameterized
by $T$. Hence, we will work in the approximation $\mu = 0$. 

At high temperature, when the electron can be considered  massless
(i.e, ultrarelativistic), the average energy of the electron is, 
$\langle E \rangle_{UR} \simeq 3.15 \;T$ \cite{kolb}.
But at lower temperatures for which the temperature-dependent electron 
mass can not be neglected,
the average electron energy $\langle E \rangle = \rho_e / n_e$ is given by
\be
\langle E \rangle = T \left(
 \int_{1/z}^{\infty} dy \frac{y^2 \sqrt{y^2 z^2 -1}}{e^y +1} \; \slash \; 
\int_{1/z}^{\infty} dy \frac{y \sqrt{y^2 z^2 -1}}{e^y +1} \right),
\ee
where $z = T/m$. But for $T \ll m$, the electron can be
considered non-relativistic $(NR)$, then 
$\langle E \rangle_{NR} \simeq  m + \frac{3}{2} T$ \cite{kolb}.
Fig. 1 shows the average electron energies $\langle E \rangle$ and
temperature dependent electron mass [see, Eq. (2)]. Here we can see that Eq. (30) 
goes to $\langle E \rangle_{UR}$ for $m_0 < T$ and becomes $\langle E \rangle_{NR}$
for $T \lsim m_0$ approximately. 
Therefore, we can use Eq. (30) for the initial electron energy
in numerical calculations of the scattering cross section. 
Fig. 2 shows the transport scattering cross section for electron-neutron scattering
$\sigma_{ne} (T)$ $(cm^{2})$
as a function of $x = T/m_0$, for the initial electron energy $E = \langle E
\rangle$ [Eq. (30)].
Fig. 3 shows the transport scattering cross sections for electron-proton scattering
$\sigma_{pe} (T)$ $(cm^{2})$
and the Coulomb scattering $\sigma_{pe}^{Coul} (T)$ $(cm^{2})$ as a
function of $x = T/m_0$ for an initial electron energy $E = \langle E \rangle$ [Eq. (30)].


\section{Conclusions}

We have calculated temperature-dependent electron-hadron transport cross sections. 
These are important, for example, in
the calculation of baryon diffusion coefficients at finite temperature.
The major motivation here has been to investigate whether finite temperature effects 
can significantly change the baryon transport cross section $\sigma_t$.
In this work, we have treated hadrons as particles which have an internal
structure and an anomalous magnetic moment. 
Also, we have assumed that their internal structure is not affected by the finite
temperature
since their mass is more than about 1000 times the temperature of interest.
 
Two major features of the finite temperature effects on the light particles
have been included in the calculation:
(1) finite temperature Dirac spinors
which are recast into the form of an effective electron mass ;
(2) finite temperature modifications to the phase space
distribution of the electrons.
We find that, for $m_0 < T$, both $\sigma_{ne}(T)$ and 
$\sigma_{pe}(T)$ approach the ultrarelativistic limit (where the electron mass can be 
ignored). In the case of electron-proton scattering, we have compared it with 
the Coulomb scattering cross section at finite temperature. In particular,
for the case of electron-neutron transport scattering, we find 
$\sigma_{ne}(T) / \sigma_{ne} (T=0) \simeq 5$ at $T \simeq 0.1 \, MeV$.

In conclusion, the baryon diffusion coefficients which affect baryon 
inhomogeneities during big-bang nucleosynthesis could be changed significantly by 
our temperature dependent electron-hadron transport cross sections.
Up to the time of weak decoupling ($T \simeq 1 \, MeV$) there is little change
in the cross sections. However, during the epoch of nucleosynthesis
($T \lsim 0.2 MeV$) when baryon diffusion is most important, 
the transport cross sections increase as the temperature decreases.
On the other hand, baryon diffusion at low temperature is strongly affected
by proton-neutron scattering for which these finite temperature effects are 
insignificant. Clearly, a study of the effects of these new cross sections
on the baryon diffusion coefficients and inhomogeneous primordial nucleosynthesis
is desired. These will be the subject of a subsequent paper.  

\vspace{0.5cm}
{\bf Acknowledgements.}

\noindent
The author (ISS) would like to thank P. Marronetti and Prof. S. Rhie
for their helpful comments. ISS also acknowledges the support by the 
Korea Research Foundation (KRF) for the Post-Doctoral Fellowship
at University of Notre Dame. This work supported in part by 
DOE Nuclear Theory Grant DE-FG02-95ER40934.


\newpage
\begin{center}
\begin{large}
FIGURE CAPTIONS
\end{large}
\vspace{5mm}\
\end{center}

\begin{itemize}
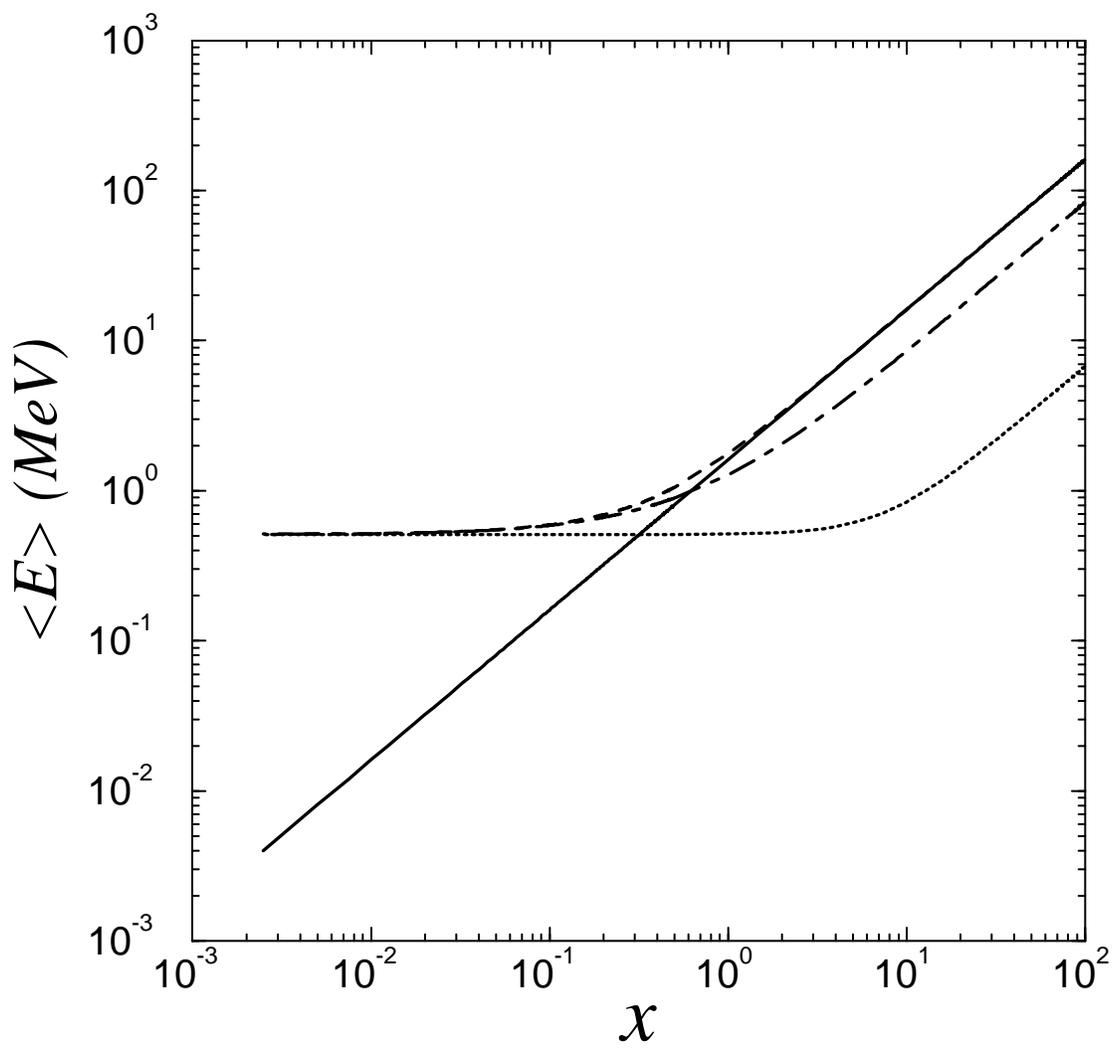
\item [{\rm FIG. 1}] 
The average electron energy $\langle E \rangle$ and
temperature-dependent effective electron mass (dot line). The solid line denotes
$\langle E \rangle_{UR}$ and the long dashed-dot line is 
$\langle E \rangle_{NR}$. The dashed line denotes 
$E = \langle E \rangle$ [Eq. (30)].

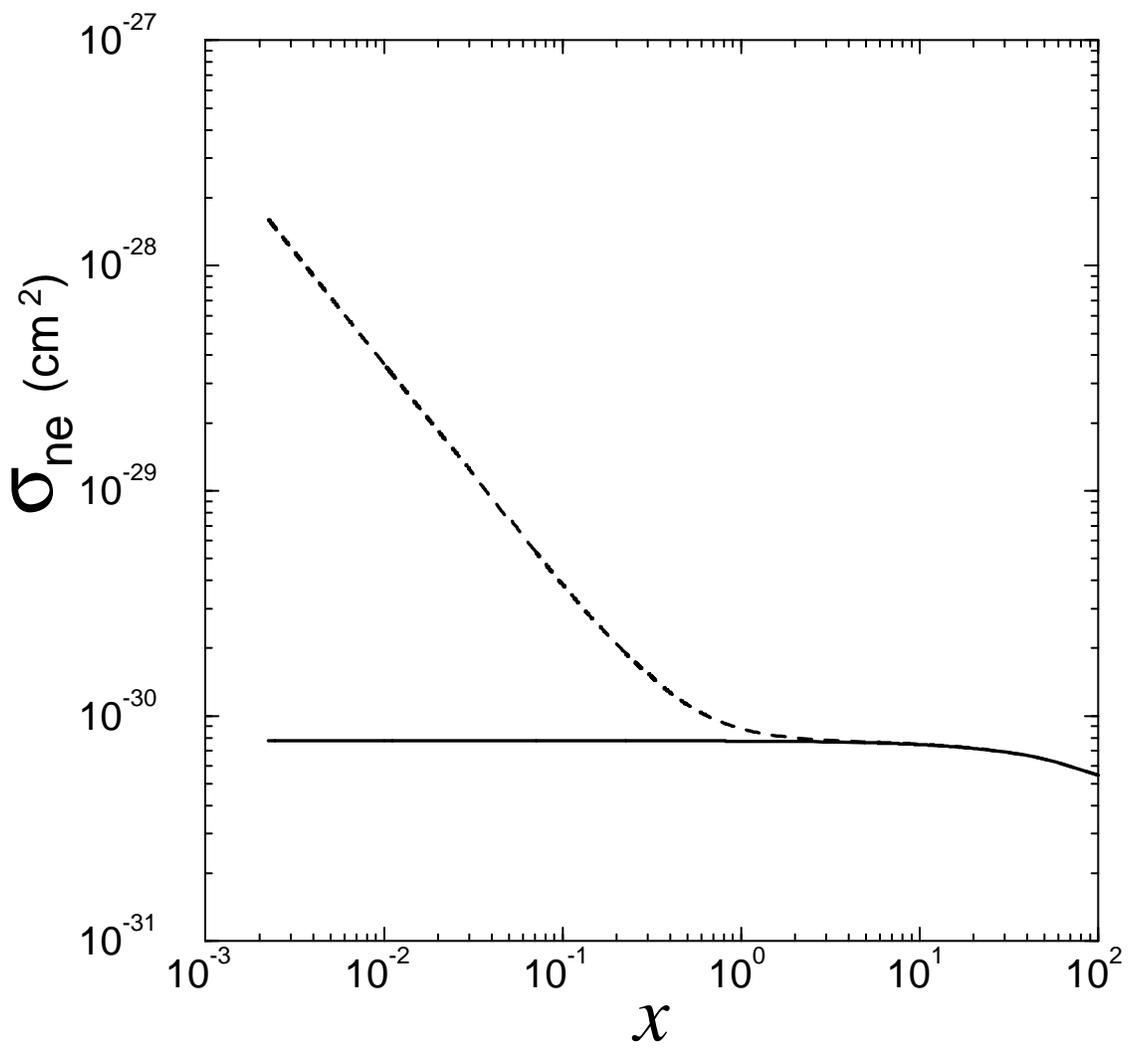
\item [{\rm FIG. 2}] 
Transport scattering cross section for electron-neutron scattering 
$\sigma_{ne} (T)$ (dashed line) compared with the ultrarelativistic approximation 
$\sigma_{ne}^{UR} (T)$ (solid line) in units of $cm^{2}$
as a function of $x = T/m_0$. This calculation assumes an initial electron 
energy $E =\langle E \rangle$ [Eq. (30)]. 

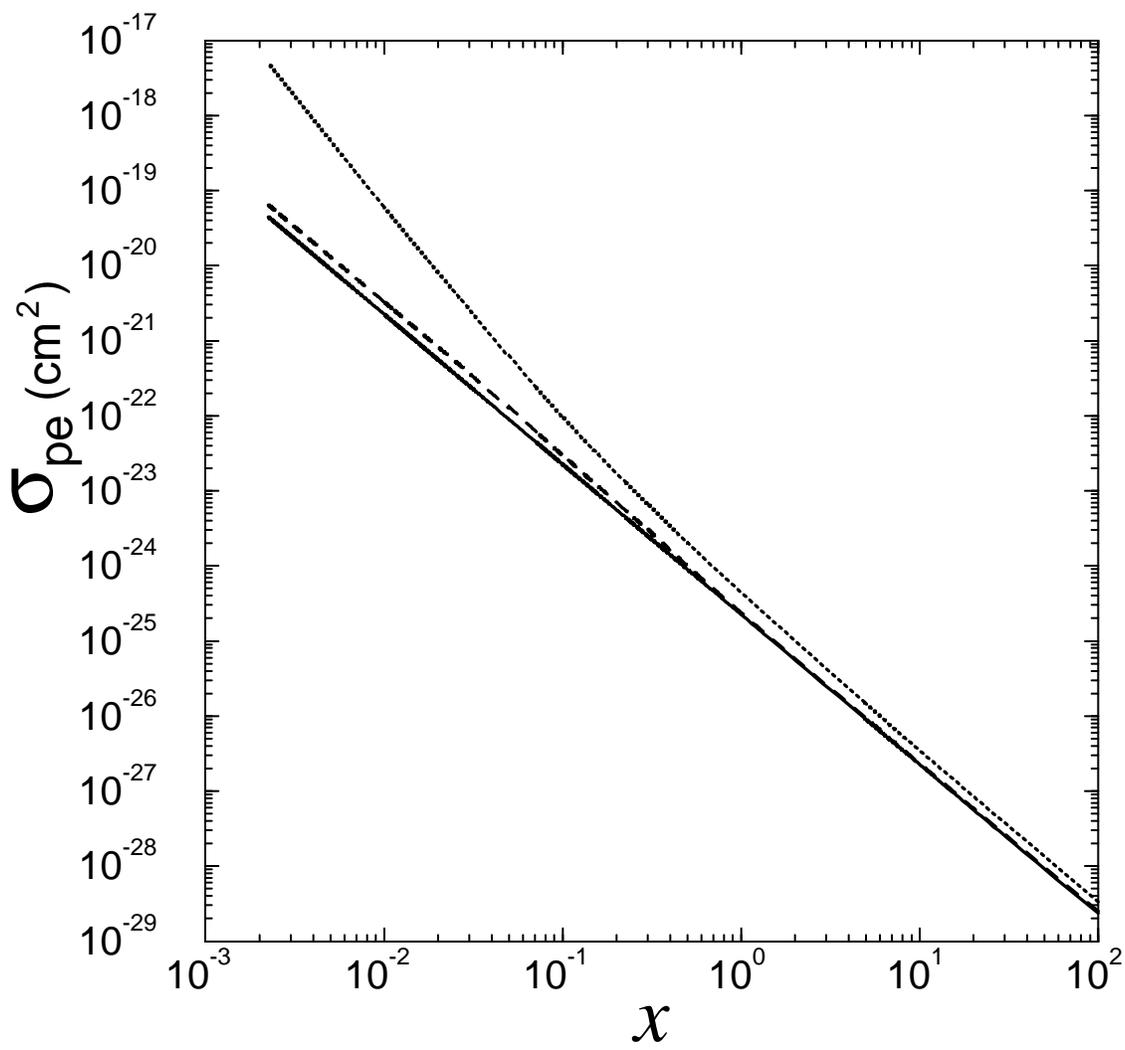
\item [{\rm FIG. 3}] 
Transport scattering cross section for electron-proton scattering
$\sigma_{pe} (T)$ (dashed line) and ultrarelativistic case 
$\sigma_{pe}^{UR} (T)$ (solid line)  
and Coulomb transport cross section $\sigma_{pe}^{Coul} (T)$ (dot line)
in units of $cm^{2}$
as a function of $x = T/m_0$, assuming an initial
electron energy $E = \langle E \rangle$ [Eq. (30)].
\end{itemize}

\end{document}